# Current density distribution in resistive fault current limiters and its effect on device stability

Farokhiyan, Mohammad[1]; Hosseini, Mehdi[1a] and Kavousi-Fard, Abdollah[2]

[1]*Department of physics, Shiraz University of Technology, Shiraz, 313-71555, Iran*

[2]*Department of electrical and electronics engineering, Shiraz University of Technology, Shiraz, 313-71555, Iran*

The increase of current uniformity along of a resistive type superconductor fault current limiter (R-SFCL) in the design of this type of limiters is well perceived as an important issue. The non-uniform distribution of current in R-SFCL only increases the current in some superconducting regions, as a result, in the fault conditions, only certain parts of the superconductor undergo a phase change that increases the heat pressure in those areas and causes the breakdown and destruction of the device. In this paper, the current density distributions in common patterns used in R-SFCs constructions have been simulated and investigated. To this end, an effective model is proposed for R-SFCL to achieve the highest uniformity of current and harmonic phase change over superconductors compared to other patterns. The simulation results in the Ansys Maxwell Software advocate the appropriate and satisfying performance of the proposed model.

# I. INTRODUCTION

In the past few decades, due to the expansion of power systems, the increasing demand and combination of local systems power supplying units in these systems, the level of fault current and short circuit have also increased, resulting in possible the instability and reduced reliability [1-5]. The appearance of varied distributed generators (DGs) ranging from renewable energy sources such as wind turbine and photovoltaics to fossil-fuel based DGs such as fuel cell and micro turbines has caused high short-circuit currents when fault happens at the distribution level. This is not only due to the internal structure of these devices; but is also due to their vicinity to the electrical load centers [6,7]. High short circuit faults impose huge costs on power system maintenance, which has made researchers to seek ways to control and limit them. Using fuses and Fault Current Limiters (FCL) are common methods for controlling the fault current [8-10]. Fault current limiters, according to their types of construction can be classified into superconducting and non-superconducting FCL [11-13]. The fuses are due to the interruption of current during a fault that results in its costs and consequences and non-superconducting FCL due to the long response time in high current, lose their performance [14]. SFCL, due to the inherent nature of the superconductor, does not have any current loss under normal conditions. In fault condition, the superconductor phase changes at a few milliseconds to a state with high resistance and reduces the fault current to lower level and after clearing the fault, it returns very fast to zero resistance state [15-17].

Depending on the type of construction and performance, SFCL's can be divided into two categories: direct and indirect SFCL [18]. The most well-known indirect SFCL can be named as inductive [19], hybrid [20] and magnetic [21]. The most important direct SFCL type is the resistive type of SFCL (R-SFCL). R-SFCL are more commonly used because of their simpler structure and smaller size [22-23]. The use of spiral and meander patterns is common to design and construct limiting fault current structures [23]. The meander patterns are also usually used for bolometers fabrication [24].

However, the main problem with this pattern is the non-uniformity of the current density along the superconductor [25]. In this paper, a method is proposed to simulate the current density distribution in different patterns used for constructing R-SFCL. The patterns are further modified through accurate simulation methods. Finally, a model is introduced that has more uniformity of current density than other patterns.

## II. SIMULATION DETAILS

Ansys Maxwell Software is used to simulate and test the current distribution in R-SFCL. This software is used to simulate both electromagnetic and electrical fields in the electromechanical and electronic systems. In these simulations, the finite element method is assessed, and by solving numerical equations, the variables are calculated [26]. The simulations are carried out for YBCO material in 2D, in the form of a thin film for the AC current, which relative data is given in Table 1 [27].

Table1. Simulated superconductor data

| | |
|---|---|
| Critical current density | $8 \times 10^6$ (A/cm$^2$) |
| Temperature | 77 (K) |
| The length of superconductor Thin Film Section | 0.5 (mm) |
| Substrate's area | 10 (mm$^2$) |
| Voltage | 40 (V) |

It is worth noting that the available methods for studying the current density distribution are mainly achieved experimentally [25, 28].

In Fig.1, some patterns that could be used to fabricate R-SFCl are shown.

In order to study the current distribution along the superconductor in our analysis, we divide the weakly superconducting regions for analyzing the current fluctuations as shown in Fig. 2. The Standard Deviation of Current Density (SDCD) for each system is calculated as equation (1) on the central path (the perpendicular bisector of dividing lines) and compared for different patterns.

$$\Gamma = \sqrt{\langle J^2 \rangle - \langle J \rangle^2} \qquad (1)$$

## III. RESULTS AND DISCUSSION

The electric current density distribution is shown in Fig. 3.

Fig. 3-a shows that for MLP the current density is approximately uniform in meander lines and is about $2.42 \times 106$ A/cm2 but in the curved regions, the electric current density is unevenly distributed so that in the sharp points current density increases up to $1.01 \times 107$ A/cm2. These regions are known as the weak superconducting regions. In SSP and CSP, the weak regions are also visible.

As it is illustrated in this figure the problems with these patterns are the uneven distribution of the electric current density along the superconductor at the sharp points (weak points) is much higher than other points. For this reason, in fault conditions, these points first undergo phase change and have higher resistance that cause increasing the thermal stress on these points, thereby causing superconductor degradation [25].

In the following the standard deviation of currents for all patterns are calculated and compared. The standard deviation of the current density is plotted for each line and along the central line (C line) in Fig. 4.

As shown in Fig 4, the maximum SDCD is related to the weak regions of this pattern, and at this voltage, phase changing is faster than what was observed in other regions of superconductor. This causes thermal pressure on weak points that makes the sample to degrade at high voltages.

To modify these patterns, the sharp areas were removed, without the change of the superconductor length so that the total resistance remained almost unchanged.

Fig 5 illustrates the current density for modified patterns. As shown in Fig. 5, sharp areas for the MMLP, MSSP and MCSP scheme are cleared away.

SDCD for modified patterns are compared with the first patterns in fig 6. This figure shows that the peaks that are clear in Fig. 4 are not observed for the modified patterns. In the MLP pattern two peaks of SDCD with an approximate value of $1/6 \times 106$ A/cm2 are visible which is not visible in the MMLP pattern and the current density is evenly distributed along the superconductor. The disappearance of the peaks can also be seen in the other two patterns. Therefore, it could be concluded that the phase change over the superconductor occurs more uniformly and the heat pressure on the weak superconductor region is reduced.

At the end, we introduce a Semicircular Meander Pattern (SMP) pattern that has the highest uniform distribution of current density as shown in Fig. 7.

Fig. 8 shows the SDCD comparison of the proposed scheme to other modified patterns that illustrates the performance of this pattern in a high voltage.

As shown in Fig. 8, the SMP pattern (newly introduced) has the most uniform distribution of current density. As a result, in this model, the superconductor becomes more uniformly phase-changing, which reduces the thermal pressure and increases the superconducting physical resistance.

Another feature of this pattern is to increase the superconductor length at a constant level compared to other patterns, which also increases its total resistance after phase change, as well as increasing dissipative energy, which leads to the increase of limited fault current.

Furthermore, the important parameter in the construction of R-SFCL is the filling factor. The filling factor is defined as the ratio of the superconducting surface to the substrate surface. The filling factor and superconductor length for all pattern is calculated and is presented in table 2.

Table2. The filling factor and superconductor length for all pattern

| Patterns/Parameters | filling factor | pattern length |
|---|---|---|
| Meander | 59.85 % | 120 mm |
| Modified meander | 57.87 % | 117 mm |
| Square spiral | 56.8 % | 105 mm |
| Modified square spiral | 54.81 % | 102 mm |
| Circular spiral | 37.73 % | 84 mm |
| Modified circular spiral | 38.29 % | 80 mm |
| Semicircular meander | 62.68% | 124 mm |

## IV. CONCLUSION

Unbalanced dispersion of the current density at high currents can lead to device degradation. In this paper, the numerical method is introduced to investigate the distribution of current density for the patterns that be used in R-SFCL fabrication.

As the results show, in order to increase the superconductivity stability of the R-SFCL that results in the stability of the power systems, it is better to use modified patterns in the construction of the R-SFCL. Comparison of the modified patterns and the first patterns shows that in the modified patterns the electric current density is more evenly distributed and in addition the filling remains almost constant. Filling factor retention also means that the superconductor resistance remains constant under fault conditions.

## V. REFERENCES

[1] Ye L. & Campbell A.M., Electric power systems research, **77**, 534–539, (2007).

[2] Fotuhi-Firuzabad, M., Aminifar, F., & Rahmati, I. IEEE Transactions on Power Delivery, **27**, 610-617, (2012).

[3] Kim, J. S., Lim, S. H., & Kim, J. C. IEEE Transactions on Applied Superconductivity, **20**, 1159-1163, (2010).

[4] Blair, S. M., Booth, C. D., Singh, N. K., Burt, G. M., & Bright, C. G. IEEE Transactions on Applied Superconductivity, **21**, 3452-3457, (2011).

[5] Firouzi, M., Gharehpetian, G. B., & Mozafari, B. Electric Power Components and Systems, **43**, 234-244, (2015).

[6] Sung, B. C., Park, D. K., Park, J. W., & Ko, T. K. IEEE transactions on industrial electronics, **56**, 2412-2419, (2009_.

[7] Perera, N., Rajapakse, A. D., & Buchholzer, T. E. IEEE transactions on power delivery, **23**, 2347-2355, (2008).

[8] Naderi, S. B., Jafari, M., & Hagh, M. T. IEEE Transactions on Industrial Electronics, **60**, 2538-2546, (2012).

[9] Teng, J. H., & Lu, C. N. IET generation, transmission & distribution, **4**, 485-494, (2010).

[10]  Hagh, M. T., & Abapour, M. IEEE Transactions on Power Electronics, (24), 613-619, (2009).

[11] Alam, M. S., Abido, M. A. Y., & El-Amin, I. Energies, **11**, 1025, (2018).

[12] Ito, D., Yoneda, E. S., Tsurunaga, K., Tada, T., Hara, T., Ohkuma, T., & Yamamoto, T. IEEE Transactions on Magnetics, **28**, 438-441, (1992).

[13]  Okedu, K. E. IET Renewable Power Generation, **10**, 1211-1219, (2016).

[14] E.M, Leung , IEEE power engineering Review **20**. 15-18, (2000).

[15] Gray, K. E., & Fowler, D. E. Journal of Applied Physics, **49**, 2546-2550, (1978).

[16] Alaraifi, S., El Moursi, M. S., & Zeineldin, H. In 2013 IEEE Grenoble Conference (pp. 1-6). IEEE. (2013).

[17] El Moursi, M. S., & Hegazy, R. IEEE Transactions on Power Systems, **28**, 140-148, (2012).

[18] Jiang, Yu, et al. 2001 Power Engineering Society Summer Meeting. Conference Proceedings (Cat. No. 01CH37262). Vol. 1. IEEE, 2001.

[19] Kozak, S., Janowski, T., Wojtasiewicz, G., Kozak, J., Kondratowicz-Kucewicz, B. and Majka, M., IEEE Transactions on Applied Superconductivity **20**, 1203-1206, (2010).

[20] Choi, H.S., Cho, Y.S. and Lim, S.H., IEEE transactions on applied superconductivity **16**: 719-722, (2006).

[21] A. Hekmati, M. Hosseini, M. Vakilian, and M. Fardmanesh. Physica C: Superconductivity **472** 39-43 2012.

[23] Ye, Y., Xiao, L., Wang, H. and Zhang, Z. In 2005 IEEE/PES Transmission & Distribution Conference & Exposition: Asia and Pacific. 1-6. IEEE, (2005).

[23] Sung, B.C., Park, D.K., Park, J.W. and Ko, T.K., IEEE transactions on industrial electronics **56**: 2412-2419, (2009).

[24] M. Hosseini, A, Moftakharzadeh, A. Kokabi, M.A, Vesaghi, H, Kinder, & M. Fardmanesh, IEEE Transactions on Applied Superconductivity, **21**, 3587-3, (2011).

[25] Kang, J. S., Lee, B. W., Park, K. B., & Oh, I. S. (2004, October). In IEEE PES Power Systems Conference and Exposition. (pp. 7-11). IEEE, (2004).

[26] https://www.ansys.com/products/electronics/ansys-maxwell

[27] Y.Q. Chen, W.B. Bian, W.H. Huang, X.N. Tang, G.Y. Zhao, L.W. Li, N. Li, Y.H. Yang, W. Huo, J.Q. Jia, C.Y. You, Sci. Rep. 6 38257, (2016).

[28] Lee, B. W., Kang, J. S., Park, K. B., Kim, H. M., & Oh, I. S. IEEE transactions on applied superconductivity, **15**, 2118-2121, (2005).

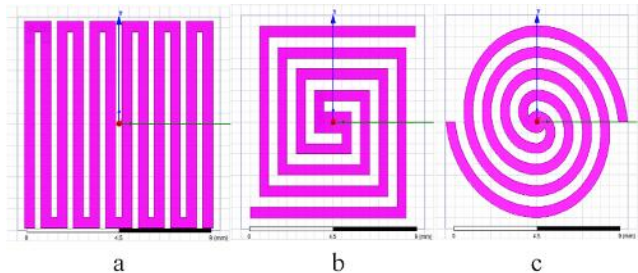

FIG. 1. Patterns used to build R-SFCL. (a) Meander Line Pattern (MLP), (b) Square Spiral Pattern (SSP) and (c) Circular Spiral Patterns (CSP).

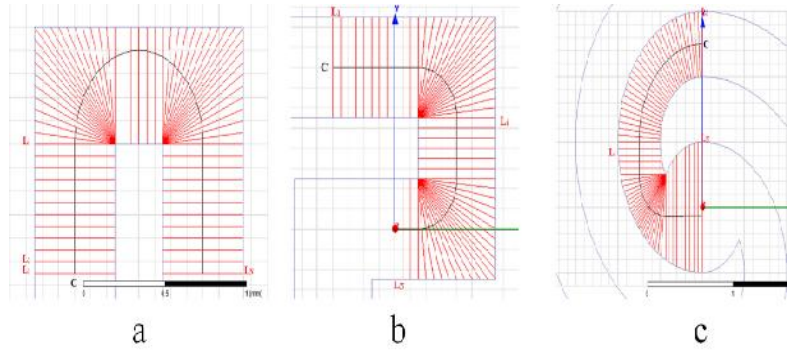

FIG. 2. The weak points and central lines of superconductor region in

(a) meander, (b) square spiral and (c) circular spiral pattern.

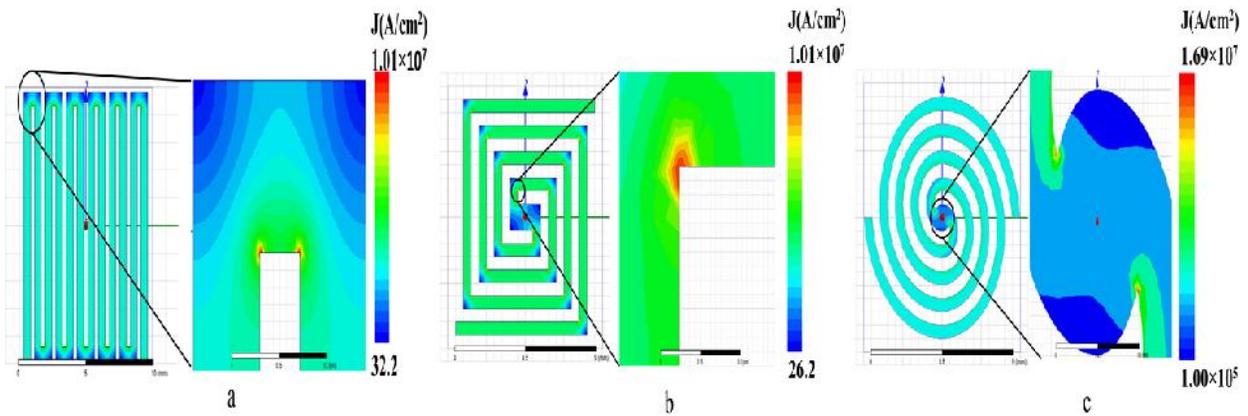

FIG. 3. Distribution of electric current density inside superconductors and weak

points in (a) MLP, (b) SSP and (c) CSP.

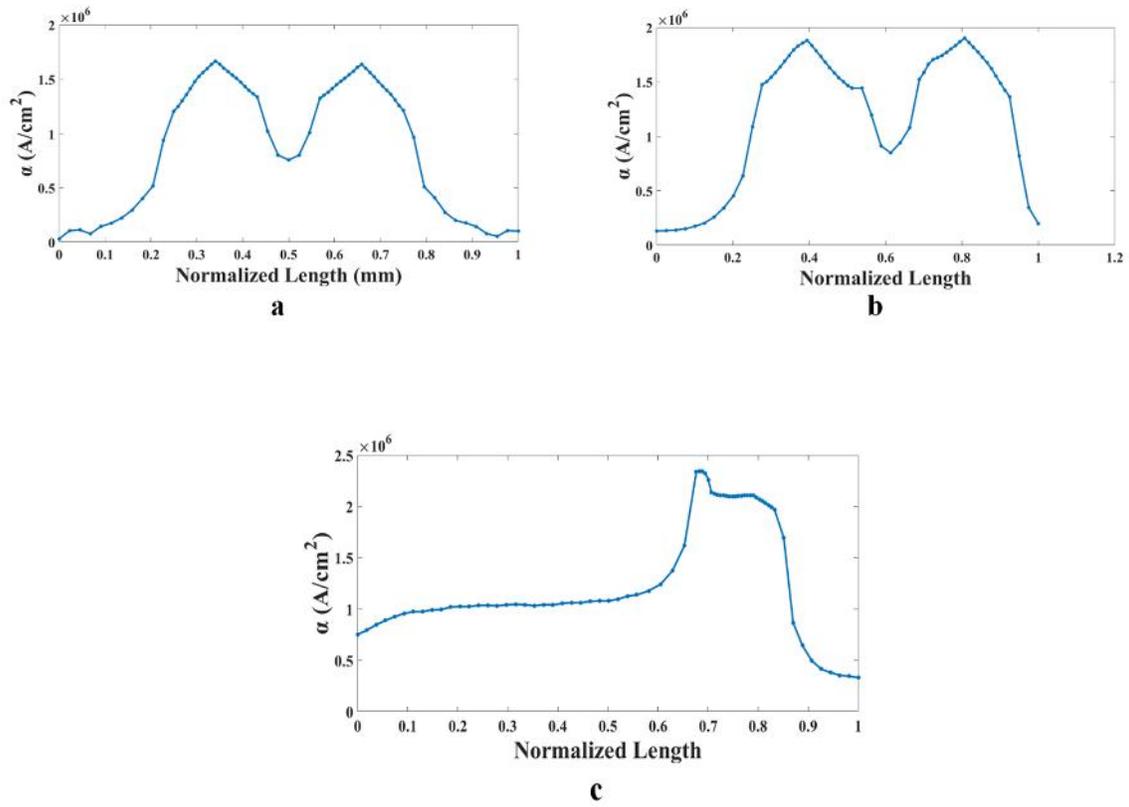

FIG. 4. SDCD of (a) MLP, (b) SSP and (c) CSP versus length of central

line of the superconductor.

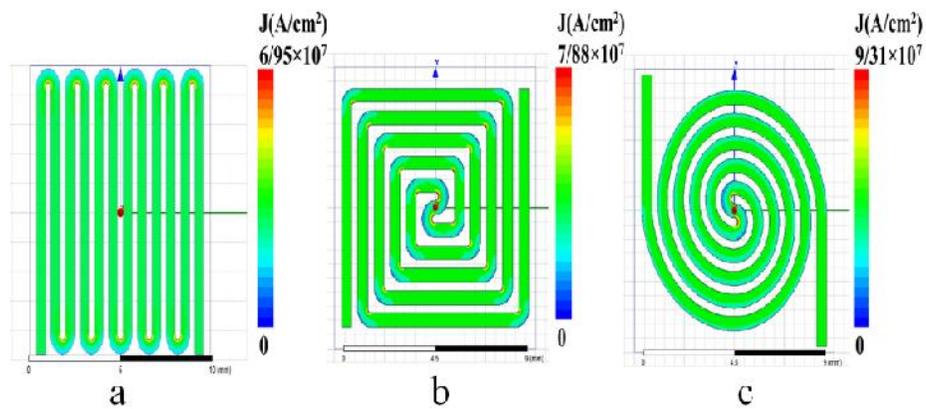

FIG. 5. (a) Modified meander line pattern (MMLP). (b) Modified square spiral pattern (MSSP). (c) Modified circular spiral pattern (MCSP).

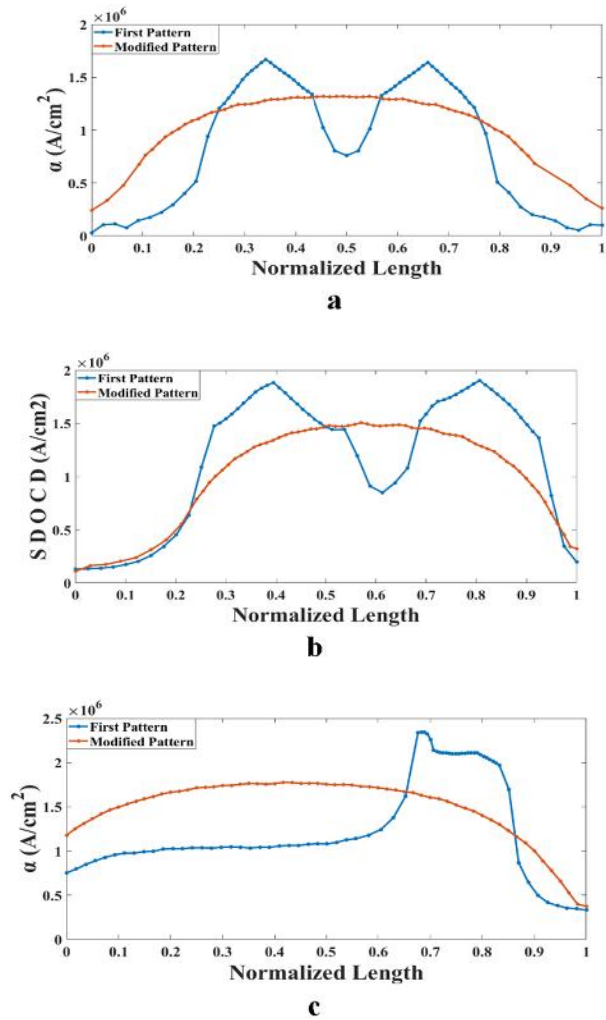

**a**

**b**

**c**

FIG. 6. Comparison of SDCD of first pattern and modified pattern for (a) MLP and (b) SSP and (c) CSP.

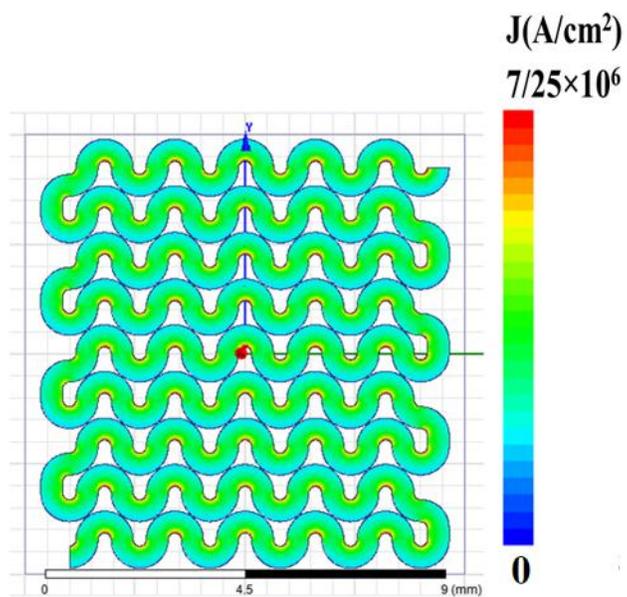

FIG. 7. Distribution of electric current density for SMP.

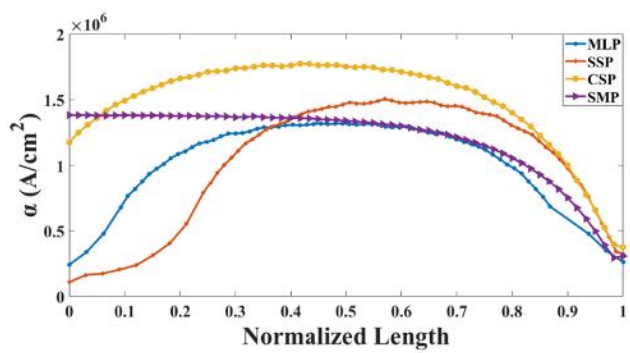

FIG. 8. SDCD Comparison of standard deviation of current density in studied patterns.